\documentclass{iopart}
\usepackage{iopams}  
\usepackage{graphicx}
\usepackage{textcomp}

\def\sh#1{#1} 

\newcommand{\CEA}{CEA-Saclay, IRAMIS, Service des Photons, Atomes et Mol\'ecules, 91191 Gif-sur-Yvette, France}
\newcommand{\PIVUT}{Photonics Institute, Vienna  University of Technology, Gu\ss hausstra\ss e 27/387, A-1040 Vienna, Austria}
\newcommand{\INRS}{Institut National de la Recherche Scientifique, Centre Energie, Mat\'eriaux et T\'el\'ecommunications, 1650 Lionel-Boulet, Varennes, Qu\'ebec J3X 1S2, Canada}
\newcommand{\moscow}{General Physics Institute of the Russian Academy of Sciences, Vavilova Street, 38, Moscow 119991, Russia}

\begin{document}

\title[]{Phase distortions of attosecond pulses produced by resonance-enhanced high harmonic generation}

\author{S. Haessler$^{1,2}$,  V. Strelkov$^3$, L. B. Elouga Bom$^4$, M. Khokhlova$^3$, O. Gobert$^1$, J.-F. Hergott$^1$, F. Lepetit$^1$, M. Perdrix$^1$, T. Ozaki$^4$, P. Sali\`eres$^1$ }
\address{$^1$\CEA}
\address{$^2$\PIVUT}
\address{$^3$\moscow} 
\address{$^4$\INRS}
\ead{\mailto{pascal.salieres@cea.fr}}

\begin{abstract}
Resonant enhancement of high harmonic generation can be obtained in plasmas containing ions with strong radiative transitions resonant with harmonic orders.
The mechanism for this enhancement is still debated. We perform the first temporal characterization of the attosecond emission from a tin plasma \sh{under near-resonant conditions for two different resonance detunings.} We show that the resonance considerably changes the relative phase of neighbouring harmonics. For very small detunings, their phase locking may even be lost, evidencing strong phase distortions in the emission process and a modified attosecond structure. These features are well reproduced by our simulations, allowing their interpretation in terms of the phase of the recombination dipole moment.
\end{abstract}

\submitto{\NJP}
\maketitle

\section{Introduction}
The strong-field process of high harmonic generation (\textsc{hhg}) provides a source of coherent femtosecond and attosecond pulses in the extreme ultraviolet (\textsc{xuv}) with a wide variety of applications \cite{Krausz2009}. It has allowed the study of photoionization of atoms and molecules with attosecond resolution \cite{Schultze2010,Kluender2011probing}, revealing unexpected features, in particular near resonances \cite{Haessler2009ion,Caillat2011molbit}. One would expect that such resonances also affect the \textsc{hhg} process itself, since the final step of the conventional semi-classical model \cite{Schafer1993,Corkum1993Plasma} is photo-recombination, \textit{i.e.} the inverse process of photo-ionization \cite{Le2009QRS,Frolov2010Potentialbarrier,Frolov2011analytic}. However, few experiments have shown evidence of resonance effects \cite{Toma1999,Sheehy1999,Woerner2009,Higuet2011cooper,Shiner2011Xe}, until recently when \textsc{hhg} was demonstrated from ablation plasmas produced on solid targets \cite{Ganeev2007}. This technique allows using a wide variety of ionic targets for \textsc{hhg}, including some with very strong radiative transitions. Enhancement of single harmonics resonant with such transitions has resulted in conversion efficiencies of $\sim10^{-4}$ \cite{Suzuki2006Anomalous,Ganeev2006Strong}, far exceeding the typical value of $\sim10^{-5}$ obtained in neutral gases \cite{Hergott2002Extreme}.

This resonance-enhancement is currently a hot topic of both applied and fundamental interest. However, so far, only the spectral intensity of high harmonics from resonant plasmas has been experimentally characterized and the important question whether they offer a route to intense XUV femtosecond or attosecond pulses remains unanswered. More fundamentally, theories describing the still scarcely explored role of resonances, and thus atomic structure, in strong-field processes \cite{Strelkov2010,Tudorovskaya2011resonance,Milosevic2007resonant,Milosevic2010,Redkin2010,Elouga2008Correlation,Kulagin2009,Frolov2010Potentialbarrier}, would greatly benefit from advanced experimental characterization of the emission as a benchmark. As yet, it was shown that the resonance-enhancement effect vanishes when shifting the harmonic energy off the resonance \cite{Suzuki2006Anomalous,Ganeev2006Strong,Ganeev2012twocolortin}. Furthermore, the xuv yield was found to decrease with increasing ellipticity of the driving laser\cite{Suzuki2006Anomalous}, suggesting electron recollision as a crucial part of the enhancement mechanism. The ellipticity dependence is, however, much slower than for rare gas media \cite{Budil1993ellipticity}.

In this article, we report measurements of both the femtosecond and attosecond structure of \textsc{hhg} from tin ablation plasma, where harmonic 17 (H17) of an \hbox{$\approx800$-nm} laser is tuned into resonance with the \hbox{4d$^{10}$5s$^2$5p $^2$P$_{3/2} \leftrightarrow$  4d$^{9}$5s$^2$5p$^2$~($^1$D)$^2$D$_{5/2}$} transition in Sn$^+$ with energy $\epsilon_\mathrm{res}=26.27\:$eV. We find a femtosecond envelope that does not correspond to a slowly-decaying plasma emission but is indeed slightly shorter than the driving laser pulse. However, the harmonic relative phase, which governs the attosecond structure, is significantly perturbed by the resonance. In resonant conditions, the phase locking between the resonant and the neighbouring orders is lost, \textit{i.e.} their relative phase varies significantly within the harmonic spectral width. We develop a model based on the numerical solution of the time-dependent Schr\"odinger equation (TDSE) that reproduces the experimentally observed strong phase perturbations, both on the attosecond and the femtosecond scales. Furthermore, we clearly link these effects to the resonance by means of an analytically calculated recombination dipole matrix element.

\section{Experimental study}
Our experimental setup is similar to the one in \cite{Elouga2011chrome}. The \mbox{20-Hz}, \mbox{50-mJ} Ti:sapphire laser is split into two beams before compression: a $\approx 300$-ps pre-pulse is loosely focused to $\sim10^{10}\:$W$\:$cm$^{-2}$ intensity onto the solid tin target at normal incidence, creating an ablation plume of $\lesssim1\:$mm length. The second beam is delayed by $80\:$ns and then compressed to $\tau_\mathrm{IR}=55\:$fs duration, before being focused ($f/50$) to $I_\mathrm{IR}^\mathrm{pump}\lesssim 1\times10^{14}$W$\:$cm$^{-2}$ (estimated by measuring the attochirp of the harmonic emission from an argon gas jet in the same experimental conditions \cite{Mairesse2003Attosecond}) into the plasma plume, propagating at a distance of $\approx 300\:$\textmu m parallel to the target surface. The annular far-field profile of the \textsc{hhg} driving beam allows blocking it with an iris $\approx1\:$m downstream, while the harmonics pass on-axis. The generated harmonics are detected by photo-ionizing argon atoms in a magnetic bottle electron spectrometer (\textsc{mbes}). Here, a weak \textsc{ir} probe beam ($I_\mathrm{IR}^\mathrm{probe}\sim10^{11}\:$W$\:$cm$^{-2}$) can be overlaid, leading to two-photon \textsc{xuv}$\pm$\textsc{ir} ionization. Thus, between the odd-order harmonic peaks, even-order sidebands (SBs) appear in the photo-electron spectrum \cite{Bouhal1997}, which serve as the nonlinear signal for the temporal characterization. In these very challenging experiments, the signal-to-noise ratio was limited by the relatively small number of pre-pulses (few thousands) that the tin target could withstand before deterioration and drop in harmonic signal \cite{Elouga2011chrome}.

\begin{figure}[tb]
\flushright
\includegraphics[width=.8\columnwidth]{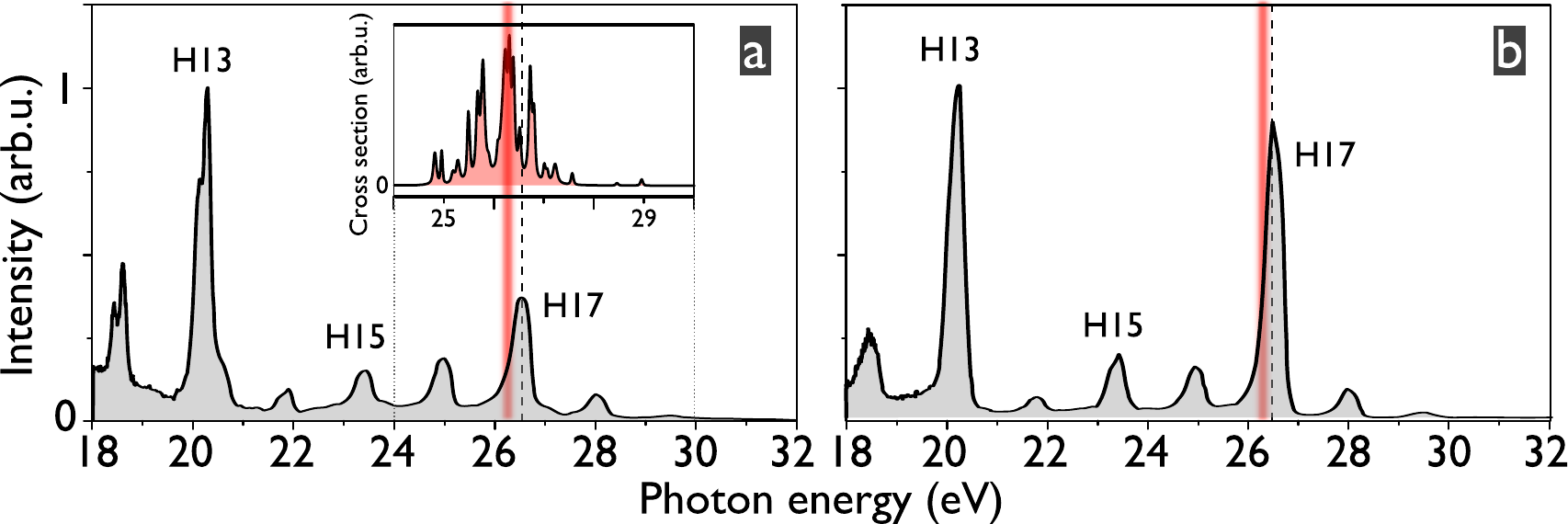}
\caption{\label{Fig:spectra} Spectra obtained by photoionizing Ar gas with high harmonics generated in tin plasma for the `detuned' (a) and `resonant' (b) cases, in presence of the \textsc{ir} probe beam. We corrected for the ionization cross-section such that the quantity shown is proportional to the \textsc{xuv} spectral intensity, except for the presence of SBs, that \emph{only} appear in the photoelectron spectrum. The $0.16\:$eV wide red line marks the resonance and the dashed line marks the exact position of H17. The inset in panel (a) shows the theoretical photoionization cross section of Sn$^+$ \cite{Ganeev2012twocolortin}, on the same energy scale as the \textsc{hhg} spectrum.}
\vspace{-12pt}
\end{figure}
Two separate measurement runs with the laser central wavelengths $\lambda_\mathrm{L}=793\:$nm and 796 nm were made, corresponding to detunings of H17 from the resonant wavelength $\lambda_\mathrm{res}=ch/\epsilon_\mathrm{res}$ of $\vert\lambda_\mathrm{res}-\lambda_\mathrm{IR}/17\vert/\lambda_\mathrm{res}
=7.5\times10^{-3}$ and $1.13\times10^{-2}$, respectively. While this difference is small, it shifts the H17 photon energy by $0.1\:$eV, which is significant compared to the $0.2\:$-eV typical width of  a harmonic line and the $\approx0.16\:$-eV resonance width \cite{Ganeev2012twocolortin}. The relative positions of H17 and the resonance for the two driving wavelengths are marked in figure \ref{Fig:spectra}: for $\lambda_\mathrm{L}=796\:$nm, H17 is closer to the exact resonance than for $793\:$nm, and for the sake of brevity, we will in the following use the oversimplified terms of \emph{`resonant'} and \emph{`detuned'} case, respectively. Note that we expect the resonance to play a role in the emission of H17 in both cases. We do not consider the AC Stark shift of $\epsilon_\mathrm{res}$ which is expected to be small at our laser intensities \cite{Ganeev2012twocolortin}. \sh{A possible small blue-shift of $\lambda_\mathrm{L}$ due to a time-dependent electron density during the driving pulse is also neglected. No such blue-shift, or other significant phase distortions have been observed in \textsc{spider} measurements of the driving laser pulses after propagation through the plasma plume.}  Note finally the array of 4d $\leftrightarrow$ 5p transitions around 26 eV (see inset in figure \ref{Fig:spectra}a). The above-mentioned transition to the ($^1$D)$^2$D$_{5/2}$ autoionizing (AI) state is expected to play the dominant role because of its \emph{gf}-value exceeding those of the others by a factor $>3.5$ \cite{Ganeev2012twocolortin}.

The distinct difference in resonance conditions is evident in the \textsc{hhg} spectra shown in figure \ref{Fig:spectra}. The laser focusing geometry, pulse energies and durations were kept the same in both runs, which ensures equal intensities for the pre-pulse and driving pulse, as well as equal phase matching conditions, assuming similar spatial beam quality. However, H17 shows significantly stronger enhancement in the `resonant' case, while H13 and H15 are of essentially the same intensities in both cases. In the `detuned' case, varying the generation conditions and in particular increasing the \textsc{hhg} driving intensity did not allow us to increase the H17 relative intensity. We thus attribute the difference to the driving laser wavelength.

The enhancement observed in figure \ref{Fig:spectra}(b) is smaller than that reported in \cite{Suzuki2006Anomalous}. This may be due to the different macroscopic plasma medium conditions, caused by our relatively large ($\approx300\:$\textmu m)  distance of the driving beam from the target (so as to avoid high electron density gradients that diffract the \textsc{ir} light, hampering the blocking of the driving beam), and smaller pre-pulse spot size resulting in a shorter generating medium. 

\subsection{Femtosecond characterization}
\begin{figure}[tb]
\flushright
\includegraphics[width=\columnwidth]{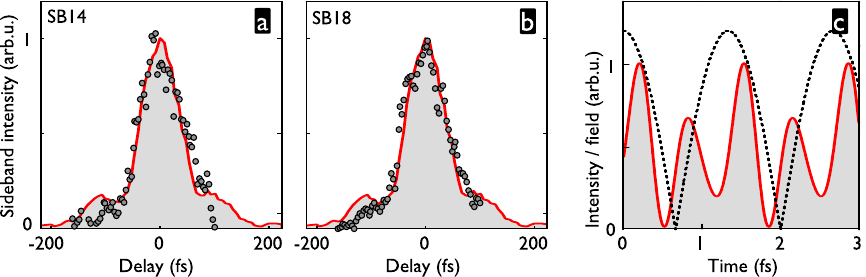}
\caption{\label{Fig:temporal} Temporal characterization of the harmonic emission from tin plasma plume. (a)/(b) Measured data (circles) with 5-point Savitzky-Golay smoothing, from low resolution \textsc{xuv}-\textsc{ir} cross-correlations for SBs 14/18 in the `resonant' case. The gray shaded area is an intensity auto-correlation of the \textsc{ir} laser pulse. (c) Attosecond pulse profile (gray shaded area), in the `detuned' case, reconstructed from the \textsc{rabbit} scan shown in figure \ref{Fig:rabitt}b,c, taking into account harmonics 13 to 19. The dashed line is the \textsc{ir} driving field modulus.}
\vspace{-12pt}
\end{figure}
In a first step, low-resolution ($3.3\:$fs delay steps) \textsc{xuv}-\textsc{ir} cross-correlation scans were performed to measure the \textsc{xuv} pulse envelopes. The SB intensity $I_\mathrm{SB}(\tau)=\int_{-\infty}^\infty I_\mathrm{\textsc{xuv}}(t) f[I_\mathrm{\textsc{ir}}(t-\tau)]\:\rmd t$ provides a cross-correlation of the \textsc{xuv} and the \textsc{ir} pulses, with temporal intensities  $I_\mathrm{\textsc{xuv}}(t)$ and  $I_\mathrm{\textsc{ir}}(t)$, respectively \cite{Bouhal1997}. The function $f[I_\mathrm{\textsc{ir}}]$ is well approximated by a power law $\propto {I_\mathrm{\textsc{ir}}}^\alpha$ at low \textsc{ir} intensity; in our experimental conditions, we find $\alpha\approx0.8$ \cite{Haessler2012femtosecondplasma}. Figures \ref{Fig:temporal}(a) and (b) show the intensities of the SBs 14 and 18 as a function of the \textsc{xuv}-\textsc{ir} delay $\tau$ in the `resonant' case. Given the high intensity contrast of H13 vs. H15 and H17 vs. H19, these two SBs essentially compare the durations of H13 and H17, $\it{i.e.}$ a non-resonant with a resonant harmonic. Within the experimental precision, both cross-correlation traces have the same shape and width  ($\tau_\mathrm{SB}=(80\pm10)\:$fs FWHM) as the intensity auto-correlation of the 55-fs laser pulse (measured using a standard second-order auto-correlator).  Data from the `detuned' case show a very similar behaviour \cite{Haessler2012femtosecondplasma}. 
The harmonic duration, $\tau_\mathrm{XUV}$, can be estimated by assuming Gaussian intensity profiles, such that $\tau_\mathrm{XUV}=({\tau_\mathrm{SB}}^2-{\tau_\mathrm{IR}}^2 /\alpha )^{1/2}$ \cite{Bouhal1997}. We then find  $\tau_\mathrm{XUV}=(50\pm15)\:$fs for both H13 and H17. For a 55-fs driving pulse, this is fairly long, especially for harmonics close to the cutoff. A likely reason is that \textsc{hhg} was saturated before the driving pulse peak. Indeed, the observed \textsc{hhg} cutoff at H19 suggest an effective driving intensity of $\gtrsim6\times10^{13}\:$W$\:$cm$^{-2}$, which is already higher than the barrier suppression intensity \cite{Ilkov1992} for the ionization of Sn$^+$ ($I_\mathrm{p}=14.6\:$eV).

\subsection{Attosecond characterization}
In order to access the attosecond sub-structure of the \textsc{hhg} emission, the \textsc{xuv}-\textsc{ir} cross-correlation scans have been repeated varying $\tau$ with sub-laser-cycle resolution (0.15-fs delay steps) over $\approx7\:$fs around the peak overlap. The \textsc{ir} probe beam intensity has been reduced by a factor $\approx2$ to completely suppress higher than first-order SBs \cite{Swoboda2009intensity}. This is an implementation of the \textsc{rabbit} technique \cite{Paul2001Observation}: 
the high resolution in $\tau$ allows observing oscillations of the SB intensities with twice the \textsc{ir} laser frequency, $\omega_\mathrm{L}$, as a result of interference between the two quantum paths involving absorption of the neighbouring harmonic photons and either absorption or emission of an \textsc{ir} photon.
From the phase of this oscillation in each SB $q$, the relative phase, $\Delta\varphi_{q}$, of two neighbouring harmonic orders ($q-1$) and ($q+1$) can be extracted. This yields an approximation to the \textsc{xuv} group delay (GD) $\left.\partial \varphi/\partial\omega \right\vert_{q\omega_\mathrm{L}}\approx \Delta\varphi_{q}/2\omega_\mathrm{L}$, where $\omega_\mathrm{L}$ is the laser frequency.

More precisely, since we are not dealing with discrete harmonic lines, one can write the $\tau$-dependent sideband intensity spectrum as $S\hspace{-.4mm}B(\omega,\tau)=\vert A(\omega+\omega_\mathrm{L})\exp[\rmi\omega_\mathrm{L}\tau] + A(\omega-\omega_\mathrm{L})\exp[-\rmi\omega_\mathrm{L}\tau] \vert^2$, where $A(\omega)$ is the harmonic complex spectral amplitude.  The integration over $\omega$ of the SB line, as usually done in experiments (the convolution with the probe-\textsc{ir} spectrum ($0.05\:$eV FWHM) and with the spectrometer resolution ($0.25\:$eV FWHM) in any case washes out the $\omega$-dependence of the $2\omega_\mathrm{L}$-oscillation \textit{within} the SB line), thus gives access to a mean value of the harmonic phase difference \emph{if the latter does not vary significantly over the harmonic spectral width:} the two harmonic orders are then said to be locked in phase. A significant variation would result in a disappearance of the $2\omega_\mathrm{L}$-oscillation: this has never been observed in any media, be it atomic gases \cite{Paul2001Observation,Mairesse2003Attosecond}, molecular gases \cite{Boutu2008Coherent,Haessler2010tomo} or non-resonant plasmas \cite{Elouga2011chrome}.   Such a disappearance would correspond to significant variations of the individual attosecond pulses throughout the emitted pulse train \cite{Varju2005,Varju2005Reconstruction}.

\begin{figure}[tb]
\flushright
\includegraphics[width=\columnwidth]{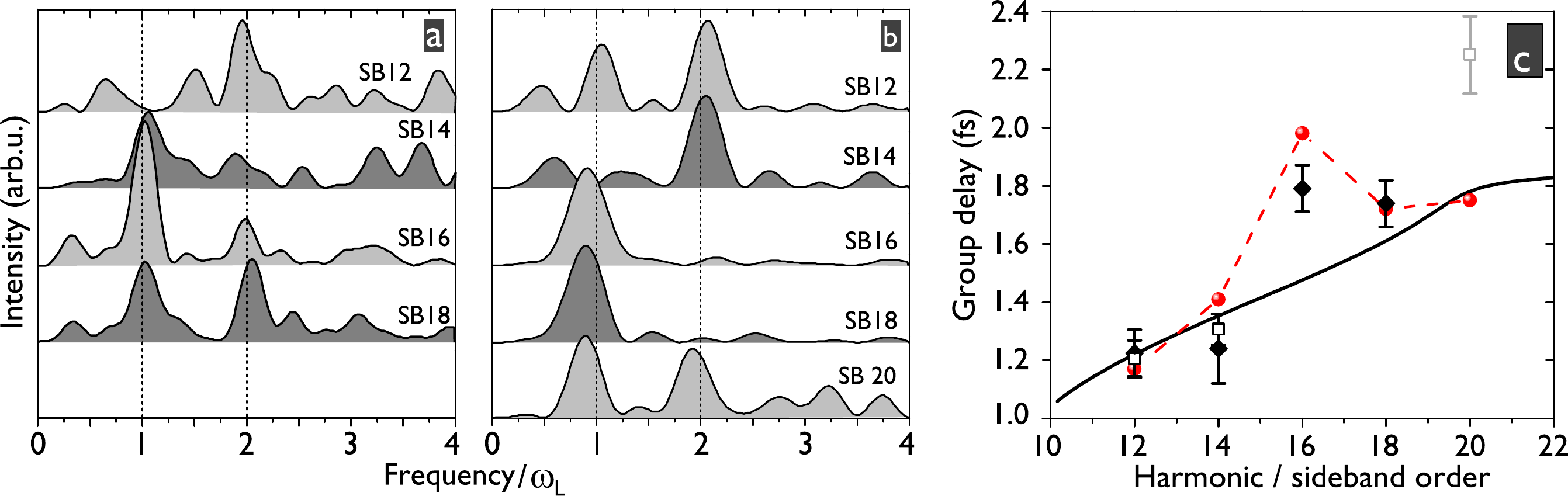}
\caption{\label{Fig:rabitt} Results of \textsc{rabbit} measurements. Squared modulus of the Fourier-transformed oscillating SB signals in the `detuned' (a) and `resonant' (b) cases. The phase at the peaks near $2\omega_\mathrm{L}$ gives the GDs, shown in (c) for the `detuned' ($\blacklozenge$) and `resonant' cases ($\opensquare$). Errors are the standard deviation of the phase within the FWHM of the $2\omega_\mathrm{L}$-peak.  Red dots and dashed line show GDs extracted from numerical simulations for the `detuned' case (see text) under a driving intensity of 6.5$\times10^{13}\:$W$\:$cm$^{-2}$. Solid line shows an SFA-calculation \cite{Mairesse2003Attosecond} for this intensity and the ionization potential of Sn$^+$.}
\vspace{-12pt}
\end{figure}

Figures \ref{Fig:rabitt}(a) and (b) show the squared Fourier transform of the $\tau$-dependent integrated SB intensities from \textsc{rabbit} scans for the  `detuned' and `resonant' case, respectively. In the `detuned' case, the $2\omega_\mathrm{L}$-oscillation is detected with reasonable contrast on all observable SBs. In the `resonant' case, we find a striking peculiarity of the two SBs (16 and 18) surrounding the resonant H17: though present in the photoelectron spectrum and of similar amplitude as in the `detuned' case (see figure \ref{Fig:spectra}), they \emph{do not} oscillate at $2\omega_\mathrm{L}$. This oscillation is detected with good contrast on all other SBs to which the resonant H17 does not contribute, demonstrating that interferometric stability was sufficient. The strong intensity difference between H17 and H15, that leads to a reduced interference contrast, is the same between H15 and H13 and does not prevent observation of the $2\omega_\mathrm{L}$-oscillations of SB14. We thus attribute this suppression to the closer proximity of H17 to the exact resonance.

From the SBs with detectable $2\omega_\mathrm{L}$-oscillation, the GDs shown in figure \ref{Fig:rabitt}(c) are obtained. For SB12 and SB14, the measured GDs in the `resonant' and `detuned' cases agree well within the experimental precision, underlining that the experimental conditions have been equal, except for the small laser wavelength shift that tunes H17 into resonance. This also shows that, despite the reduced signal-to-noise ratio, leading sometimes to small shifts of the $2\omega_\mathrm{L}$-peak positions, the measured phases are reliable. The GDs at SB12 and SB14 also agree well with those predicted by the strong field approximation (SFA) for the short trajectory class and a driving intensity of  $6.5\times10^{13}\:$W$\:$cm$^{-2}$ (solid line in figure \ref{Fig:rabitt}(c)). For SB16 and SB18, we can only extract GDs for the `detuned' case. For SB16, we find a striking increase of the GD by $350\:$as as compared to the SFA prediction. As we will show, this strong phase distortion is a result of the nearby resonance. For SB18, the measured GD is much closer to the SFA prediction again, even though a big instability for this specific SB was observed in the different data sets.
Finally, for SB20 we only obtain a good SB oscillation contrast in the `resonant' case and for a single \textsc{rabbit} scan. The extracted GD carries a rather large error due to the low signal-to-noise ratio and may not be reliable. We will thus not discuss it further and show it only for completeness.

In figure \ref{Fig:temporal}(c), we show the average attosecond pulses in the train reconstructed through a Fourier transform from the measured GDs and intensities of harmonics 13 to 19 in the `detuned' case. The rather big attochirp together with the phase jump between H15 and H17 (corresponding to the GD shift of SB16) distorts the pulse shape to a double peak per half cycle.

\section{Theoretical study and discussion}
\begin{figure}[tb]
\centering
\includegraphics[width=.5\columnwidth]{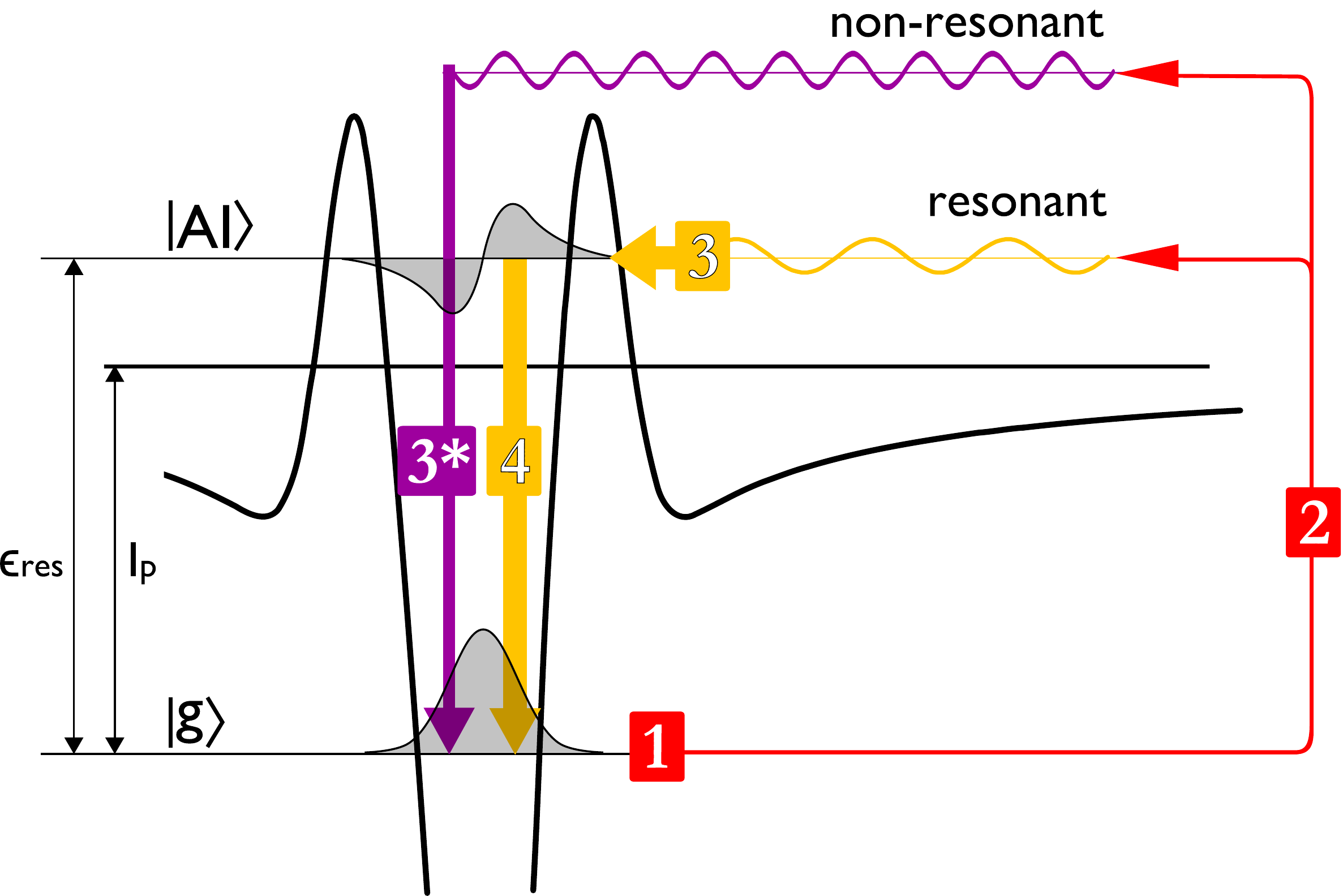}
\caption{\label{Fig:4steps} Illustration of the role of the AI state in resonant HHG. See text for details.}
\vspace{-12pt}
\end{figure}

In order to interpret our experimental observations, we made simulations  based on the numerical solution of the 3D-TDSE for a single active electron in a model ionic potential, modulated by a strong laser field. The potential (similar to the one shown as a black line in figure \ref{Fig:4steps}; see details in~\cite{Strelkov2010,Ganeev2012twocolortin}) has a quasi-stationary state modeling the ($^1$D)$^2$D$_{5/2}$ AI state that we expect to dominate the resonance-enhancement effect. Namely, we reproduce the ionization and the AI state energy, the width and (less accurately) the oscillator strength of the ground--AI state transition in real Sn$^+$. The laser pulse has a $\cos^2$-intensity-envelope with a duration of $55\:$fs FWHM. 
The computed time dependent expectation value of the dipole acceleration gives the harmonic spectrum in amplitude and phase. 

In order to partly simulate the macroscopic effects, and in particular to select the shortest quantum path contribution, we perform a coherent summation of the numerical spectra over the laser intensity (steps of 2x10$^{12}$ W/cm$^2$), thus calculating an on-axis XUV signal from a thin target~\cite{Platonenko1998hhgreview,Balcou1997phasematching}. In this signal, the shortest trajectory contribution dominates~\cite{Platonenko1998hhgreview} because phases of longer trajectories contributions vary rapidly with the laser intensity~\cite{Balcou1997phasematching}.

As it was suggested in~\cite{Strelkov2010} and further studied in~\cite{Ganeev2012twocolortin,Tudorovskaya2011resonance}, the role of the resonance with the AI state can be understood within the four-step HHG model \sh{illustrated in} figure~\ref{Fig:4steps}. After (tunnel) ionization (step 1) and acceleration in the laser field (step 2), the electron can recombine to the ground state and emit \textsc{xuv} (step $3^{*}$ in the three-step model~\cite{Schafer1993,Corkum1993Plasma}); however, if the energy of the \sh{returning} electron is close to that of the AI state, the electron can be captured \sh{such that the} system finds itself in the AI state (step 3) and \sh{only} then relaxes to the ground state, emitting \textsc{xuv} (step 4). Thus, the phase of the resonant harmonic is defined by the phase accumulated during the free-electronic motion \emph{and} during recapturing of the electron. The calculation results presented below show that the latter is very sensitive to the detuning from the resonance.   

Group delays were found from the numerical spectrum by taking the phases at the exact harmonic frequencies, $(2q\pm1)\omega_\mathrm{L}$ and calculating the phase differences
$\Delta\varphi_{q}$ and thus the GDs $\Delta\varphi_{q}/2\omega_\mathrm{L}$.
The robustness of the extraction of GDs from the TDSE solutions has been verified by comparing to two additional methods:  (ii) After selecting in the spectral domain only two neighbouring harmonics, $(2q\pm1)$, we make an inverse Fourier-transform back to the time domain, where we obtain a train of attosecond pulses. The temporal position of the first attosecond pulse peak after the driving pulse maximum gives the GD at SB $q$. (iii) Selecting in the spectral domain only single harmonic peaks,
inverse-Fourier-transforming to the time domain and multiplying by $\exp[\rmi q w_\mathrm{L} t]$, we find the q-th harmonic complex slowly varying envelope, $a_q(t)$. The harmonic temporal intensity and phase are then $\vert a_q(t)\vert^2$ and $\arg[a_q(t)]$, respectively. We then calculate the GDs from the time averaged relative phases of neighbouring harmonics. The results of these three methods agree to within $\pm60\:$as.

In figure \ref{Fig:rabitt}c, we show the so obtained GDs in the `detuned' case for a laser peak intensity of $I_0=6.5\times10^{13}\:$W~cm$^{-2}$. Very satisfactory agreement is obtained with the measured values. In particular at SB16, the obtained GD shift of $\approx500\:$as with respect to the SFA reproduces well the experimental one. 
This shift is quite robust in the simulations for different peak intensities, $I_0$, as well as in all our measurements. In contrast, the GD at SB18 turns out to be very sensitive to $I_0$ in the simulations and for some values, large shifts are obtained. We have also observed such shifts for SB18 in a few measurements (not shown). 

The contrast of the $2\omega_\mathrm{L}$-oscillations of the different SBs was evaluated by integrating the above expression of $S\hspace{-.4mm}B(\omega,\tau)$ over the spectral width. In the `resonant' case, the contrast values for SB12 to SB20 are respectively: 0.69, 0.62, 0.08, 0.13, 0.45. In qualitative agreement with the experiments, the calculated contrast for SBs 16 and 18 is strongly reduced.  Moreover, it is significantly lower in the resonant than in the detuned case, while being approximately the same for the other sidebands.

\begin{figure}[tb]
\centering
\includegraphics[width=.5\columnwidth]{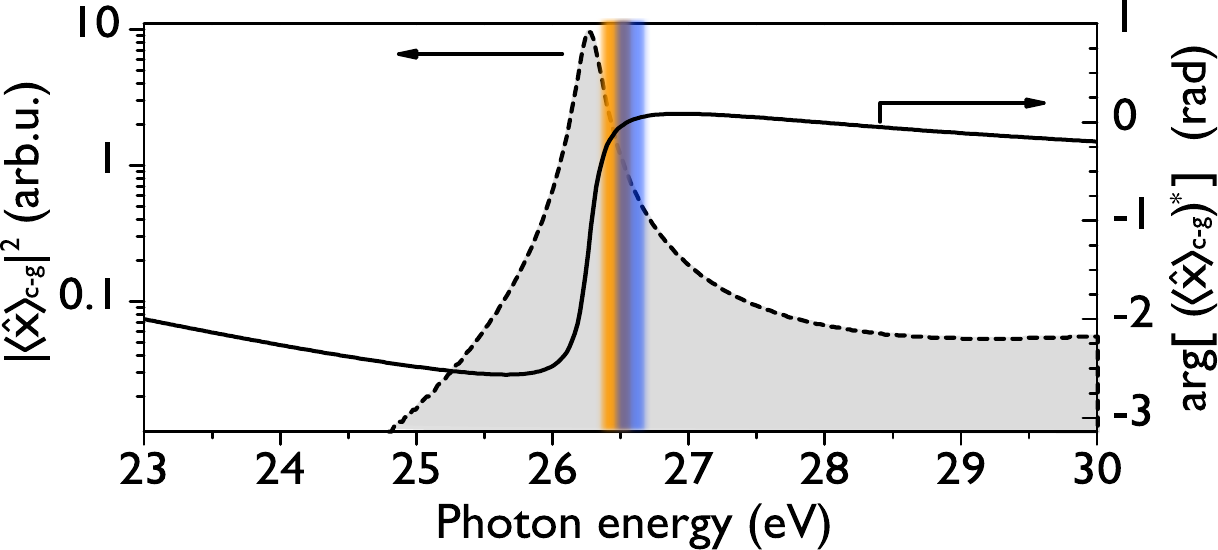}
\caption{\label{Fig:me} Phase and squared modulus of the recombination dipole matrix element $\langle \hat{x}\rangle_\mathrm{c-g}$ as a function of the emitted \textsc{XUV} photon energy (i.e. the difference of the continuum- and ground-state energy). The $0.2\:$eV wide blue and yellow lines mark the spectral position and approximate width of H17 for the detuned and resonant case, respectively. H15 and H19 are around $23.4\:$eV and $29.6\:$eV, respectively.}
\vspace{-12pt}
\end{figure}
The above results can be understood by deriving analytically the wavefunctions (both in the discrete and continuous spectrum) in a 1D double-barrier well potential in the absence of laser field. Details will be reported elsewhere \cite{Khokhlova2011poster}. With these solutions, we calculate the dipole matrix element, $\langle\hat{x}\rangle_\mathrm{c-g}$, of the continuum--ground-state transition, which plays a central role in the resonant HHG emission \cite{Le2009QRS,Strelkov2010,Frolov2010Potentialbarrier,Frolov2011analytic}. The phase of this matrix element, shown in figure \ref{Fig:me}, jumps rapidly by $2.65\:$rad at the exact resonance position, $\epsilon_\mathrm{res}=26.27\:$eV. In both the `detuned' and `resonant' cases, H17 is still on the ``blue'' side of this jump, whereas H15 is on the ``red'' side. The recombination matrix element thus adds an additional phase difference of $\Delta\varphi_{16}^\mathrm{c-g}\approx2.2\:$rad between H15 and H17, and thus shifts the GD at SB16 by $\Delta\varphi_{16}^\mathrm{c-g}/2\omega_\mathrm{L}=460\:$as. This is in very satisfactory agreement with the GD-shifts observed in the experiment and the numerical model.
Note that the dispersion of the HHG medium around the resonance leads to a phase distortion with sharp excursions of opposite sign on either side of the resonance, qualitatively different from the step-like phase induced by $\langle\hat{x}\rangle_\mathrm{c-g}$. If this propagation-induced phase were important, the corresponding group delay shift at SB16 would have to be accompanied by a similar shift \emph{with opposite sign} at SB18. Our experimental GDs are thus compatible with $\langle\hat{x}\rangle_\mathrm{c-g}$ being the dominant source of phase distortions.

The matrix element phase can also qualitatively explain the suppression of the \textsc{rabbit}-oscillations in the `resonant' case. As visible in figure \ref{Fig:me}, in the `detuned' case, it varies very little within the width of H17, while in the `resonant' case, the onset of the rapid phase jump causes a much stronger phase variation within the width of H17. This will contribute to the observed suppression of the \textsc{rabbit}-oscillation contrast on the sidebands surrounding H17. The first order of the phase variation (linear component) corresponds to a delay of H17 with respect to the other harmonics. For the `resonant' case, the delay predicted by $\langle\hat{x}\rangle_\mathrm{c-g}$ is $1.5\:$fs, which agrees well with the numerical TDSE solution. Unfortunately, this small value cannot be resolved in our experimental cross-correlations. With H17 exactly on resonance ($\lambda_\mathrm{L}=802\:$nm), $\langle\hat{x}\rangle_\mathrm{c-g}$ predicts a delay of $8\:$fs, which would certainly be measurable.

Other effects could also contribute to the reduction of the contrast of the \textsc{rabbit}-oscillation. 
For example, the resonant H17 wavefront could be spatially distorted, which would average out the relative phases of harmonic and probe beams in the \textsc{mbes} sensitive volume. Such a distortion, affecting only the resonant harmonic, could be caused by the dispersion around the resonance frequency if there were strong gradients of the Sn$^+$-density in the \textsc{hhg} medium.  But, as discussed above, the propagation-induced phase distortions seem negligible in our generation conditions due presumably to low plasma densities.

\section{Conclusion}
In conclusion, our study gives an affirmative answer to the practical question of whether resonance-enhanced HHG is indeed a source of intense ultrashort XUV pulses. The  enhanced harmonic order has the same femtosecond duration as the non-resonant ones. On the attosecond time-scale however, we find significant distortions of the phase of the near-resonant harmonic. Our results suggest the detuning from the resonance as an effective handle controlling the resonant harmonic phase.

From a more fundamental viewpoint, previous studies of the HHG phase properties focused mainly on the phase accumulated by the quasi-free electron in the continuum, or on the recombination step as a probe of molecular structure and dynamics \cite{Haessler2011tutorial,Salieres2012tomoreview}. This article presents experimental evidence of the dramatic influence of the recombination step on the phase of resonant harmonics from an atomic target.

\sh{
We have shown that the recombination dipole matrix element alone can describe the origin of the phase distortions observed in our experiments. This confirms the four-step picture \cite{Strelkov2010} illustrated in figure \ref{Fig:4steps}, where the resonance comes into play only as an intermediate state in recombination, but not as an initial state as in the model proposed in \cite{Milosevic2007resonant}. It is possible to factor out this recombination dipole matrix element from the expression for the radiating dipole in the description of HHG \cite{Le2009QRS,Frolov2011analytic}. This is the basis for the so called `self-probing' schemes to extract structural and dynamic information about the generating system from intensity, phase and polarization measurement of high harmonics \cite{Haessler2011tutorial,Salieres2012tomoreview}. Our results thus suggest the possibility of devising `self-probing' schemes for atomic resonances based on the advanced characterization of resonance-enhanced high harmonics. In particular, the rapid phase variation responsible for the intriguing suppression of the \textsc{rabitt} oscillations may encode characteristic features of the involved AI state. To gain further insight, the phase within the spectral width of the resonant harmonic could be measured by \textsc{xfrog} \cite{Mauritsson2004xfrog}, \textsc{xuv-spider} \cite{Mairesse2005HHSPIDER} or the \textsc{frog-crab} \cite{Mairesse2005} method, and the resonant harmonic could be spatially characterized ~\cite{Gautier2008}.
}

\ack
We thank M. Lein and D. Milo\v{s}evi\'{c} for fruitful discussions  and T. Auguste for SFA calculations. We acknowledge financial support through the European grants EU-FP7-IEF-MUSCULAR and EU-FP7-ATTOFEL, the French ANR-09-BLAN-0031-01, the Minist\`ere du d\'eveloppement \'economique, de l'innovation et de l'exportation du Qu\'ebec, the Russian Foundation for Basic Research, the Presidential Council on Grants of the Russian Federation (MD-6596.2012.2), as well as the `Dynasty' foundation.

\section*{References}
\bibliographystyle{iopart-num}
\bibliography{mightmightys}

\end{document}